\global\def\draftcontrol{0}

   \def\versionno{ unismall}

\catcode`\@=11

\expandafter\ifx\csname draftcontrol\endcsname\relax\global\def\draftcontrol{0}
\fi

{\count255=\time\divide\count255 by 60
\xdef\hourmin{\number\count255}
\multiply\count255 by-60\advance\count255 by\time
\xdef\hourmin{\hourmin:\ifnum\count255<10 0\fi\the\count255}}
\def\draftdate{\number\month/\number\day/\number\year\ \ \ \hourmin }

\newcommand\makepapertitle{\par
  \begingroup
    \renewcommand\thefootnote{\@fnsymbol\c@footnote}%
    \def\@makefnmark{\rlap{\@textsuperscript{\normalfont\@thefnmark}}}%
    \long\def\@makefntext##1{\parindent 1em\noindent
            \hb@xt@1.8em{%
                \hss\@textsuperscript{\normalfont\@thefnmark}}##1}%
     \newpage
     \global\@topnum\z@   
     \@makepapertitle
     \thispagestyle{empty}\@thanks
  \endgroup
  \setcounter{footnote}{0}%
  \global\let\thanks\relax
  \global\let\makepapertitle\relax
  \global\let\@makepapertitle\relax
  \global\let\@thanks\@empty
  \global\let\@author\@empty
  \global\let\@date\@empty
  \global\let\@title\@empty
  \global\let\title\relax
  \global\let\author\relax
  \global\let\date\relax
  \global\let\and\relax
  \def\version{\let\version\@version\@gobble}
}
\def\@makepapertitle{%
  \newpage
   \ifnum\draftcontrol=1 {}
   \version\versionno
   \vskip 3em%
   \else
   \hfill\hbox to 3cm {\parbox{4cm}{\@pubnum}\hss}%
   \vskip 3em%
   \fi
   \begin{center}%
   \let \footnote \thanks
     {\LARGE {\@title}}%
     \vskip 1.5em%
     {\normalsize
       \lineskip .5em%
       \begin{tabular}[t]{c}%
         \@author
       \end{tabular}\par}%
     \vskip 1.5em%
     {\@bstract}%
     \end{center}%
     \vskip 1.5em
     \@date%
   \par
}

\gdef\@pubnum{}
\def\pubnum#1{%
  \gdef\@pubnum{#1}}

\gdef\@bstract{}
\def\Abstract#1{%
  \gdef\@bstract{%
   \parbox{\textwidth-0pc}{%
   \centerline{\bf Abstract}\penalty1000%
\kern.2cm%
\noindent
\renewcommand\baselinestretch{1.0}%
{#1}}}
}

\def\ps@paper{\let\@mkboth\@gobbletwo%
     \ifnum\draftcontrol=1
    \def\@oddfoot{\hbox to \textwidth{\tiny \versionno \hfil\tiny\draftdate}%
    \hskip -\textwidth \hbox to \textwidth{\hfil\rm\thepage\hfil}}%
     \else\def\@oddfoot{\hbox to \textwidth{\hfil\rm\thepage\hfil}}
     \fi
     \let\@evenfoot\@oddfoot
}

\def\body{\clearpage
          \pagestyle{paper}
    }

\def\@version#1{\ifnum\draftcontrol=1
\typeout{}\typeout{#1}\typeout{}
\vskip3mm\centerline{\hbox{\fbox{\normalsize{\tt DRAFT -- #1 -- }
                   {\draftdate}}}}\vskip3mm
\fi}
\let\version\@version
\long\def\eqlabel#1{\ifnum\draftcontrol=1
                    \tag@false  
                    \tag*{(\theequation) \hbox to -0.2cm{\hspace{0cm}\small{#1}\hss}}
                    \refstepcounter{equation}
                    \edef\@currentlabel{\theequation}
                    \ltx@label{#1}          
                    \else
                    \label{#1}
                    \fi
                    }
\let\st@bibitem\@bibitem
\let\st@lbibitem\@lbibitem
\ifnum\draftcontrol=1
  \def\@bibitem#1{%
    \st@bibitem{#1}\a@@label{#1}\ignorespaces}
  \def\@lbibitem[#1]#2{%
    \st@lbibitem[#1]{#2}\a@@label{#2}\ignorespaces}
  \def\a@@label#1{%
    \gdef\a@lab{\smash{\normalfont\small#1}}
    \ifvmode
      \if@inlabel
        \global\setbox\@labels\hbox{%
          \llap{\a@lab\let\a@lab\relax
                \kern\@totalleftmargin\kern\marginparsep}%
          \box\@labels}%
      \fi
    \fi}
\fi

\documentclass[12pt,letterpaper]{article}

\usepackage{amsmath,amssymb,array,calc,epsfig,rotating,bm}
\usepackage[sort]{cite}
\usepackage{graphicx}
\usepackage{psfrag,verbatim}

\ifnum\draftcontrol=1
\fi

\renewcommand\baselinestretch{1.25}
\renewcommand\section{\@startsection {section}{1}{\z@}%
                                   {-3.5ex \@plus -1ex \@minus -.2ex}%
                                   {2.3ex \@plus.2ex}%
                                   {\normalfont\large\bfseries}}
\renewcommand\subsection{\@startsection{subsection}{2}{\z@}%
                                   {-3.25ex\@plus -1ex \@minus -.2ex}%
                                   {1.5ex \@plus .2ex}%
                                   {\normalfont\normalsize\bfseries}}
\renewcommand\subsubsection{\@startsection{subsubsection}{3}{\z@}%
                                   {-3.25ex\@plus -1ex \@minus -.2ex}%
                                   {1.5ex \@plus .2ex}%
                                   {\normalfont\normalsize\it}}
\renewcommand\paragraph{\@startsection{paragraph}{4}{\z@}%
                                   {-3.25ex\@plus -1ex \@minus -.2ex}%
                                   {1.5ex \@plus .2ex}%
                                   {\normalfont\normalsize\bf}}

\numberwithin{equation}{section}

\def\revise#1       {\raisebox{-0em}{\rule{3pt}{1em}}%
                     \marginpar{\raisebox{.5em}{\vrule width3pt\
                     \vrule width0pt height 0pt depth0.5em
                     \hbox to 0cm{\hspace{0cm}{%
                     \parbox[t]{4em}{\raggedright\footnotesize{#1}}}\hss}}}}

\newcommand\nxt[1]  {\\\fnxt#1}
\newcommand{\ie}{{\it i.e.,}\ }

\def\calm         {{\cal M}}
\def\caln         {{\cal N}}
\def\calo         {{\cal O}}

\def\calv         {{\cal V}}

\def\zet          {{\mathbb Z}}

\def\del          {\partial}

\def\Im           {{\rm Im\hskip0.1em}}

\def\sqr#1#2{{\vcenter{\vbox{\hrule height.#2pt
 \hbox{\vrule width.#2pt height#1pt \kern#1pt
 \vrule width.#2pt}\hrule height.#2pt}}}}
\def\square{%
  \mathop{\mathchoice{\sqr{12}{15}}{\sqr{9}{12}}{\sqr{6.3}{9}}{\sqr{4.5}{9}}}}

\newcommand{\ft}[2]{{\textstyle{\frac{#1}{#2}}}}

\def\a{\alpha}
\def\b{\beta}
\def\w{\omega}
\def\r{\rho}
\def\dd{\delta}
\def\e{\epsilon}

\def\g{\gamma}

\def\aa1{\phi}
\def\cc1{\psi}

\def\l{\lambda}

\catcode`\@=12

\begin{document}


\title{\bf Universality of small black hole instability in AdS/CFT}

\date{September 24, 2015}

\author{
Alex Buchel \\[0.4cm]
\it Department of Applied Mathematics, Department of Physics and Astronomy, \\
\it University of Western Ontario\\
\it London, Ontario N6A 5B7, Canada;\\
\it Perimeter Institute for Theoretical Physics\\
\it Waterloo, Ontario N2J 2W9, Canada
}

\Abstract{$AdS_5$ type IIb supergravity compactifications on five-dimensional
Einstein manifolds ${\cal V}_5$ realize holographic duals to
four-dimensional conformal field theories. Black holes in such
geometries are dual to thermal states in these CFTs. When black holes
become sufficiently small in (global) $AdS_5$, they are expected to
suffer Gregory-Laflamme instability with respect to localization on
${\cal V}_5$. Previously, the instability was demonstrated for
gravitational dual of ${\cal N}=4$ SYM, where ${\cal V}_5=S^5$.  We
extend stability analysis to arbitrary ${\cal V}_5$.  We point out
that the quasinormal mode equation governing the instabilities is
universal. The precise onset of the instability is ${\cal
V}_5$-sensitive, as it is governed by the lowest non-vanishing
eigenvalue $\lambda_{min}$ of its Laplacian.
}

\makepapertitle

\body

\version\versionno
\tableofcontents

\section{Introduction}\label{intro}

Consider type IIb supergravity compactification on a five-dimensional Einstein manifold $\calv_5$ 
with large five-form flux through it. The vacuum of the resulting five-dimensional effective 
gravitational action is $\calm_5=AdS_5$, dual to a strongly coupled four-dimensional conformal gauge theory
living on the boundary $\del\calm_5$. The best studied  realization of the holography is 
when $\calv_5$ is a five-dimensional sphere, $S^5$,  in which case the corresponding 
conformal gauge theory is $\caln=4$ supersymmetric Yang-Mills \cite{m1}. Other generalizations 
include the orbifolds $S^5/\zet_k$ \cite{Kachru:1998ys}, $T^{1,1}=(SU(2)\times SU(2))/U(1)$ \cite{Klebanov:1999tb} 
and $Y^{p,q}$ Sasaki-Einstein spaces \cite{Gauntlett:2004yd}. While  the details of the dual CFTs depend on 
what $\calv_5$ is chosen ( {\it e.g}, the central charge of the CFT is $\propto \frac{1}{{\rm vol}(\calv_5)}$ ),
many aspects of the theories are in fact universal. The reason for this commonality stems from the fact that 
Kaluza-Klein reduction of type IIb supergravity on $\calv_5$ contains a  {\it universal} consistently 
truncated gravitational sector\footnote{Without loss of generality 
we set the asymptotic $AdS_5$ radius to unity.}
\begin{equation}
S_5=\int_{\calm_5} d^5\xi \sqrt{-g}\left(R_5+12\right)\,.
\eqlabel{5d}
\end{equation}
We consider the case when $\del\calm_5=R\times S^3$.
Besides (global) $AdS_5$ vacuum solution, \eqref{5d} contains black holes solutions, dual to thermal states of the 
boundary CFT. In full ten-dimensional supergravity these black holes are ``smeared'' on $\calv_5$. 
The size of the black hole $\r_+$ (as measured by the radius of $S^3$ at the 
horizon\footnote{See \cite{Buchel:2015gxa} for further details and conventions. })
is related to the black hole mass $M$, compare to the vacuum energy $E_{vacuum}$ as
\begin{equation}
\r_+^2=\frac{1}{2}
\left(\sqrt{1+\epsilon}-1\right)\,,\qquad {\rm where}\qquad \e\equiv \frac{M}{E_{vacuum}}\,.
\eqlabel{rpdef}
\end{equation}
Notice that small black holes are light. It is expected that an $AdS_5$ black hole can not become arbitrarily small: 
 it was proposed in 
\cite{Banks:1998dd,Horowitz:1999uv} that in the limit $\r_+\to 0$ 
it would suffer a Gregory-Laflamme (GL) instability  \cite{Gregory:1993vy}, resulting in its localization on $\calv_5$. 
The latter localization phenomenon was explicitly verified in  \cite{Hubeny:2002xn,Dias:2015pda,Buchel:2015gxa}
when $\calv_5=S^5$.  

One might expect that small $AdS_5$ black hole localization would depend on details of $\calv_5$. The purpose of this paper 
is to explicitly demonstrate that this is not the case --- all what matters for determining the onset of the instability
is the smallest non-vanishing eigenvalue of the scalar Laplacian on $\calv_5$, feeding into the fluctuation equation 
originally constructed by Prestidge in \cite{Prestidge:1999uq}, and later 
obtained for the case  $\calv_5=S^5$ in  \cite{Hubeny:2002xn}.  

The rest of the paper is organized as follows. In section \ref{sec2} we derive a single "master'' second-order 
quasinormal mode equation in a radial $AdS_5$ coordinate.
Besides $\r_+$, this equation depends parametrically only on the quasinormal mode frequency 
$\omega$ and the scalar Laplacian on $\calv_5$ eigenvalue $\lambda$.   
It is precisely the  quasinormal mode  equation eq.(5.3) obtained in \cite{Buchel:2015gxa}
for the case $\calv_5=S^5$. For the quasinormal mode at the threshold of instability, 
\ie for $\w=0$, this equation reduces to the threshold equation of Prestidge  \cite{Prestidge:1999uq}.

\section{Stability of $AdS_5$ black holes smeared on $\calv_5$}\label{sec2}

Holographic dual to thermal states of a large class of  conformal gauge theories on $R\times S^3$ is described by
the following type IIb supergravity background
\begin{equation}
\begin{split}
&ds_{10}^2=(d\calm_5)_{BH}^2+(d\calv_5)^2\,,\qquad F_5={\rm vol}_{\calm_5}-{\rm vol}_{\calv_5}\,,\\
&F_5=\star_{10}\ F_5\,,\qquad dF_5=0\,.
\end{split}
\eqlabel{background}
\end{equation} 
where $(d\calm_5)_{BH}^2$ is the global $AdS_5$ Schwarzschild black hole metric, 
\begin{equation}
\begin{split}
&(d\calm_5)_{BH}^2 = g_{\mu\nu}dx^\mu dx^\nu=-c_1(x)^2\ dt^2 + c_2(x)^2\ dx^2 + c_3(x)^2\ (dS^3)^2\,,\\
& c_1=\frac{\sqrt{a(x)}}{\sqrt{x}}\,,\qquad c_2=\frac{1}{2x\sqrt{1-x}\sqrt{a(x)}}\,,\qquad c_3=\frac{\sqrt{1-x}}{\sqrt{x}}\,,\\
& a=\frac{(x_h+x(1-x_h))(x_h-x)}{x_h^2 (1-x)}\,,\qquad x_h=\frac{1}{1+\r_+^2}\,,\qquad x\in (0,x_h)\,,
\end{split}
\eqlabel{5dmetric} 
\end{equation}
and 
\begin{equation}
(d\calv_5)^2=g_{\a\b} dy^\a dy^\b\,.
\eqlabel{vmetric}
\end{equation}
$(d\calv_5)^2$  is the metric on an Einstein manifold $\calv_5$, and $(dS^3)^2$ is a round metric on 
unit radius $S^3$. We use $\mu,\nu,\r,\cdots $ indices on $\calm_5$, and $\a,\b,\g,\cdots $
indices on $\calv_5$. 

We are interested in $SO(4)$-invariant linearized fluctuations of \eqref{background} that carry an arbitrary angular momentum on $\calv_5$. Generically, 
metric fluctuations would couple with the fluctuations of the 5-form $F_5$ \cite{Buchel:2015gxa}. 
Substantial simplification can be achieved with the judicious choice of the gauge. Let 
\begin{equation}
\begin{split}
&\dd g_{\mu\nu}=h_{\mu\nu}(t,x,y^\a)\equiv\biggl\{\dd g_{tt},\dd g_{xx},\dd g_{tx}, \dd g_{ij}\equiv g_{ij}(x)\ \dd f(t,x,y^\a) \biggr\}\,,\\ 
&\dd g_{\mu\a}=h_{\mu\a}(t,x,y^\b)\,,\qquad \dd g_{\a\b}=h_{\a\b}(t,x,y^\g)\,,
\end{split}
\eqlabel{metricfluc}
\end{equation}
where we explicitly enumerated non-vanishing components of $\dd g_{\mu\nu}$ consistent with $SO(4)$ symmetry --- $i,j$ are angles on $S^3$.
To leading order in metric fluctuation, the linearized components of the ten-dimensional Ricci tensor on $\calv_5$ 
take form \cite{DeWolfe:2001nz}
\begin{equation}
\begin{split}
R_{\alpha\beta}^{\;\; (1)} =& -{1\over 2} \biggl[ \left(\square_{\calm_5} + \square_{\calv_5}\right) h_{(\alpha \beta)} - 2 R_{\alpha \gamma \delta \beta} h^{(\gamma \delta)} - R_\alpha^{\;\; \gamma} h_{(\gamma \beta)} - R_\beta^{\;\; \gamma} h_{(\gamma \alpha)} \\
&+ \frac15 g_{\alpha \beta} \left(\square_{\calm_5} + \square_{\calv_5}\right) h^\gamma_\gamma
- \frac{16}{15} \nabla_\alpha \nabla_\beta h^\gamma_\gamma 
+ \nabla_\alpha \nabla_\beta \biggl(h^\mu_\mu +\frac 53 h_\g^\g\biggr)\\
& - \nabla_\alpha \nabla^\mu h_{\mu \beta} 
- \nabla_\beta \nabla^\mu h_{\mu \alpha}\biggr]\,, 
\end{split}
\eqlabel{riccicompact}
\end{equation}
where 
\begin{equation}
h_{\a\b}=h_{(\a\b)}+\frac 15 g_{\a\b} h^{\g}_{\g}\,.
\eqlabel{trace}
\end{equation}
Assuming that 
\begin{equation}
h_\mu^\mu=0\,,\qquad h_{\mu\a}=h_{\a\b}=0\,,
\eqlabel{gaugechoice}
\end{equation}
we see that 
\begin{equation}
R_{\alpha\beta}^{\;\; (1)} =0\,.
\eqlabel{const1}
\end{equation}
Notice that with \eqref{gaugechoice}, 
\begin{equation}
\dd {\rm vol}_{\calm_5}\propto \calo(h^2)\,,\qquad \dd  {\rm vol}_{\calv_5}=0 \,,
\eqlabel{const2}
\end{equation}
which implies that it is consistent, at  order $\calo(h)$, with the 5-form self-duality constraint and
its Bianchi identity  not to deform the background 5-form ansatz \eqref{background}.  The latter implies that the 
5-form stress-energy tensor with components on $\calv_{5}$ is unchanged from its background value, \ie
\begin{equation}
T_{\alpha\beta}^{\;\; (1)} =0\,.
\eqlabel{const3}
\end{equation}
Together, \eqref{const1} and \eqref{const3} imply that \eqref{gaugechoice} solves all Einstein equations with indices on $\calv_5$.

For linearized components of the Ricci tensor with mixed indices we have  \cite{DeWolfe:2001nz}
\begin{equation}
\begin{split}
R_{\mu\alpha}^{\;\; (1)} =& -{1\over 2} \biggl[ \square_{\calm_5} h_{\mu\alpha} -
\nabla_\mu \nabla^\nu h_{\nu \alpha} - R_{\mu}^{\;\; \nu} h_{\nu
\alpha} + \square_{\calv_5} h_{\mu \alpha} - R_\alpha^{\;\; \beta} h_{\beta
\mu} \\ 
&- \nabla_\alpha \nabla^\nu h_{\nu\mu} + \nabla_\mu \nabla_\alpha
\left(h^\rho_\rho +h^\gamma_\gamma\right) - \nabla_\mu
\nabla^\beta h_{\beta \alpha} \biggr] \,.
\end{split}
\eqlabel{mixed}
\end{equation} 
Since in a gauge \eqref{gaugechoice} the mixed components of the 5-form stress-energy tensor vanish, 
\begin{equation}
T_{\mu\alpha}^{\;\; (1)} =0\,,
\eqlabel{const4}
\end{equation}
the Einstein equations with mixed indices reduce to 
\begin{equation}
 \nabla_\alpha \nabla^\nu h_{\nu\mu}=0\,.
\eqlabel{mixedeoms}
\end{equation} 
Parameterizing the (traceless) fluctuations as 
\begin{equation}
\begin{split}
&\dd g_{tt}= -c_1(x)^2\ e^{-i\w t}\ f_1(x)\ Y_{\calv_5}(y^\a)\,,\qquad \dd g_{xx}= c_2(x)^2\ e^{-i\w t}\ f_2(x)\ Y_{\calv_5}(y^\a)\,,\\
&\dd g_{tx}= i\ e^{-i\w t}\ f(x)\ Y_{\calv_5}(y^\a)\,,\qquad \dd g_{ij}= g_{ij}(x)\ e^{-i\w t}\ \biggl( -\frac 13 f_1(x)-\frac 13 f_2(x)\biggr)\ Y_{\calv_5}(y^\a)\,,
\end{split}
\eqlabel{expfluc}
\end{equation}
equations \eqref{mixedeoms} are equivalent to the following two equations
\begin{equation}
\begin{split}
&0=\biggl( 2\w f_1 c_2 -f \left(\ln\frac{f c_1 c_3^3}{c_2}\right)' \biggr)\ \del_\a  Y_{\calv_5}(y^\g)\ e^{-i\w t}\,,\\
&0=\biggl( f_2\ \left(\ln {f_2 c_1 c_3^4}\right)'+f_1 \left(\ln \frac{c_3}{c_1}\right)'-\frac{\w f}{2 c_1^2} \biggr)\ \del_\a  
Y_{\calv_5}(y^\g)\ e^{-i\w t}\,.
\end{split}
\eqlabel{riccimixed}
\end{equation}
 
Remaining Einstein equations\footnote{We used here equations of motion for the background \eqref{background}. } 
involve components $ _{tt}\,,\, _{tx}\,,\,  _{xx} $ and a pair of the $S^3$ components (by $SO(4)$ symmetry):
\nxt $ _{tt}$ components:
\begin{equation}
\begin{split}
0=& \biggl[ \biggl\{
\frac{c_1^2}{c_2^2}\left(f_1''+f_1' \left(\ln\frac{c_1c_3^3}{c_2}\right)'-2f_2' (\ln c_1)' \right)+\frac{\w f}{c_2^2} \left(\ln \frac{f c_3^3}{c_2}\right)'
-8 f_2 c_1^2 -\w^2 f_1
\biggr\}\ Y_{\calv_5} \\
&+f_1 c_1^2\ \square_{\calv_5} Y_{\calv_5} \biggr]\ e^{-i \w t}\,;
\end{split}
\eqlabel{eq1}
\end{equation}   
\nxt $ _{tx}$ components:
\begin{equation}
\begin{split}
0=&i \biggl[ \w \biggl\{
f_1 \left(\ln\frac{f_1 c_3}{c_1}\right)'+f_2 \left(\ln\frac{f_2 c_3^4}{c_1}\right)'
\biggr\}\ Y_{\calv_5}+\frac 12 f\  \square_{\calv_5} Y_{\calv_5}
\biggr]\ e^{-i \w t}\,;
\end{split}
\eqlabel{eq2}
\end{equation}   
\nxt $ _{xx}$ components:
\begin{equation}
\begin{split}
0=&\biggl[ \biggl\{ f_2''+
f_2' \left(\ln\frac{c_1c_3^5}{c_2}\right)'+2 f_1' \left(\ln\frac{c_3}{c_1}\right)'-\frac{\w f}{c_1^2}\left(\ln\frac{f}{c_2}\right)'
+f_2 c_2^2\left(8-\frac{\w^2}{c_1^2}\right)
\biggr\}\ Y_{\calv_5}\\
&- f_2 c_2^2\  \square_{\calv_5} Y_{\calv_5}
\biggr]\ e^{-i \w t}\,;
\end{split}
\eqlabel{eq3}
\end{equation}   
\nxt $ _{S^3S^3}$ components:
\begin{equation}
\begin{split}
0=&\biggl[ \frac 13 \biggl\{ \frac{c_3^2}{c_2^2}\biggl( f_2''+
f_2' \left(\ln\frac{c_1c_3^9}{c_2}\right)'+f_1''+f_1' \left(\ln\frac{c_1c_3^3}{c_2}\right)'\biggr) +\frac{2}{c_2^2}\left(c_3^2\right)' \left(\ln c_1 c_3\right)'
\ (f_1+4 f_2)\\&- \frac{3\w f (c_3^2)'}{2c_1^2 c_2^2}
-(f_1+f_2) c_3^2\left(8-\frac{\w^2}{c_1^2}\right)
\biggr\}\ Y_{\calv_5}
+ \left(\frac 13f_1+ \frac13 f_2\right) c_3^2\  \square_{\calv_5} Y_{\calv_5}
\biggr]\ e^{-i \w t}\,.
\end{split}
\eqlabel{eq4}
\end{equation}   
When $Y_{\calv_5}$ is an eigenfunction on $\calv_5$, \ie 
\begin{equation}
\square_{\calv_5} Y_{\calv_5}=-\lambda\ Y_{\calv_5}\,,
\eqlabel{deflambda}
\end{equation}
the PDEs \eqref{riccimixed}-\eqref{eq4} reduce to a  system of ODEs in $AdS_5$  radial coordinate $x$. 
The resulting system of ODEs is over-determined, which can be exploited to reduce it to the following equations:
\begin{equation}
\begin{split}
0=&2 c_3 \w \biggl( c_2^2 c_1^2 \lambda+4 c_2^2 c_1^2-c_2^2 \w^2-(c_1')^2 \biggr)\ f_2+2 \w \biggl( 4 c_2^2 c_1^2 c_3-c_2^2 c_3 \w^2-3 c_1 c_1' c_3'\\
&-c_3 (c_1')^2 \biggr)\ f_1
+\biggl( c_1^2 c_3' \lambda-c_1 c_3 c_1' \lambda+3 c_3' \w^2 \biggr)\ f\,,\\
0=&f'+f\ \left(\ln \frac{c_1c_3^3}{c_2}\right)'-2 \w c_2^2\ f_1\,,\\
0=&f_1'+\left(\frac{\w}{2  c_1^2} -\frac{\l}{2\w}\right)\ f-2 (\ln c_1)'\ f_2\,.  
\end{split}
\eqlabel{finals}
\end{equation}
Remarkably, a solution to \eqref{finals}, together with \eqref{deflambda}, solves \eqref{riccimixed}-\eqref{eq4}.

Using the first two equations we can algebraically eliminate $f_1$ and $f_2$ in favor of $\{f,f'\}$. 
The third equation in \eqref{finals} produces the "master'' quasinormal mode equation:
\begin{equation}
\begin{split}
&0=f''+f'\  \biggl(3 (\r_+^2+1)^3 y^{14}-(5 \r_+^4+5 \r_+^2+9) (\r_+^2+1)^2 y^{12}-\r_+^2 (\r_+^2+1)^2 (\l+22) y^{10}\\
&-\r_+^2 (\r_+^2+1) (9 \l \r_+^4+61 \r_+^4+9 \l \r_+^2+3 \w^2+61 \r_+^2+2 \l+18) y^8+\r_+^2 (-11 \w^2 \r_+^4\\
&+8 \l \r_+^4-11 \w^2 \r_+^2+11 \r_+^4+8 \l \r_+^2-3 \w^2+11 \r_+^2+3 \l+12) y^6+\r_+^4 (10 \l \r_+^4+21 \r_+^4\\
&+10 \l \r_+^2-4 \w^2+21 \r_+^2+5 \l+19) y^4+\r_+^6 (-\w^2+\l+4) y^2-\r_+^8 (\l+3)\biggr)\biggl(\\
&y (y^2-1) (\r_+^2+y^2) ((\r_+^2+1) y^2+\r_+^2) 
(-(\r_+^2+1)^2 y^8-\r_+^2 (\r_+^2+1) (\l+6) y^4\\
&+\r_+^2 (-\w^2+\l+4) y^2+\r_+^4 (\l+3))\biggr)^{-1}
+\biggl(4 (\r_+^2+1)^4 y^{20}\\
&-(8 (2 \r_+^4+2 \r_+^2+3)) (\r_+^2+1)^3 y^{18}-(\r_+^2+1)^2 (-4 \r_+^8-8 \r_+^6+7 \l \r_+^4+44 \r_+^4+7 \l \r_+^2\\
&+48 \r_+^2-24) y^{16}-\r_+^2 (\r_+^2+1)^2 (22 \l \r_+^4+96 \r_+^4+22 \l \r_+^2+\w^2+96 \r_+^2-13 \l-112) y^{14}\\
&+\r_+^2 (\r_+^2+1) (9 \l \r_+^8+72 \r_+^8+18 \l \r_+^6+\l^2 \r_+^4-2 \w^2 \r_+^4+144 \r_+^6+76 \l \r_+^4+\l^2 \r_+^2\\
&-2 \w^2 \r_+^2+416 \r_+^4+67 \l \r_+^2+6 \w^2+344 \r_+^2-6 \l) y^{12}+\r_+^4 (\r_+^2+1) (2 \l^2 \r_+^4+23 \w^2 \r_+^4\\
&+39 \l \r_+^4+2 \l^2 \r_+^2+23 \w^2 \r_+^2+184 \r_+^4+2 \l \w^2+39 \l \r_+^2-2 \l^2+42 \w^2+184 \r_+^2-58 \l\\
&-120) y^{10}+\r_+^4 (\l^2 \r_+^8-15 \l \r_+^8+2 \l^2 \r_+^6-48 \r_+^8+4 \l \w^2 \r_+^4-30 \l \r_+^6-5 \l^2 \r_+^4+42 \w^2 \r_+^4\\
&-96 \r_+^6+4 \l \w^2 \r_+^2-114 \l \r_+^4-6 \l^2 \r_+^2+\w^4+42 \w^2 \r_+^2-240 \r_+^4-2 \l \w^2-99 \l \r_+^2+\l^2\\
&-14 \w^2-192 \r_+^2+14 \l+40) y^8-\r_+^6 (-2 \l \w^2 \r_+^4+6 \l^2 \r_+^4-6 \w^2 \r_+^4-2 \l \w^2 \r_+^2+48 \l \r_+^4\\
&+6 \l^2 \r_+^2-2 \w^4-6 \w^2 \r_+^2+48 \r_+^4+6 \l \w^2+48 \l \r_+^2-4 \l^2+37 \w^2+48 \r_+^2-49 \l-128) y^6\\
&-\r_+^8 (2 \l^2 \r_+^4+\l \r_+^4+2 \l^2 \r_+^2-\w^4-24 \r_+^4+6 \l \w^2+\l \r_+^2-6 \l^2+32 \w^2-24 \r_+^2-63 \l\\
&-148) y^4+\r_+^{10} (-2 \l \w^2+4 \l^2-9 \w^2+35 \l+72) y^2+\r_+^{12} (\l+4) (\l+3)\biggr)
\biggl(\\
&(y^2-1)^2 ((\r_+^2+1)^2 y^8+\r_+^2 (\r_+^2+1) (\l+6) y^4-\r_+^2 (-\w^2+\l+4) y^2\\
&-\r_+^4 (\l+3)) ((\r_+^2+1) y^2+\r_+^2)^2 (\r_+^2+y^2)^2 y^2\biggr)^{-1}\ f\,,
\end{split}
\eqlabel{master}
\end{equation} 
where we introduced a new radial coordinate $y$, so that 
\begin{equation}
x\equiv \frac{y^2}{y^2+\r_+^2}\,,\qquad y\in (0,1)\,.
\eqlabel{defyx}
\end{equation}
This is our universal quasinormal mode equation: the only information about $\calv_5$
is in the choice of the scalar Laplacian eigenvalue $\l$.  

Equation \eqref{master} is identical to eq.(5.3) derived in \cite{Buchel:2015gxa}, provided we identify\footnote{
We refer the reader to \cite{Buchel:2015gxa} for the details associated with solving \eqref{master}.}
\begin{equation}
f_{xy}\ \Longrightarrow\ \frac{y}{y^2+\r_+^2}\ f\,,\qquad s\ \Longrightarrow\ \l\,.
\eqlabel{idbl}
\end{equation}
Additionally, when $\w=0$ it reduces to the equation at the threshold of instability, originally derived in  \cite{Prestidge:1999uq}.

\begin{figure}[t]
\begin{center}
\psfrag{x}{{$\r_+$}}
\psfrag{y}{{$-\frac{\Im\w}{2\pi T}$}}
\includegraphics[width=4in]{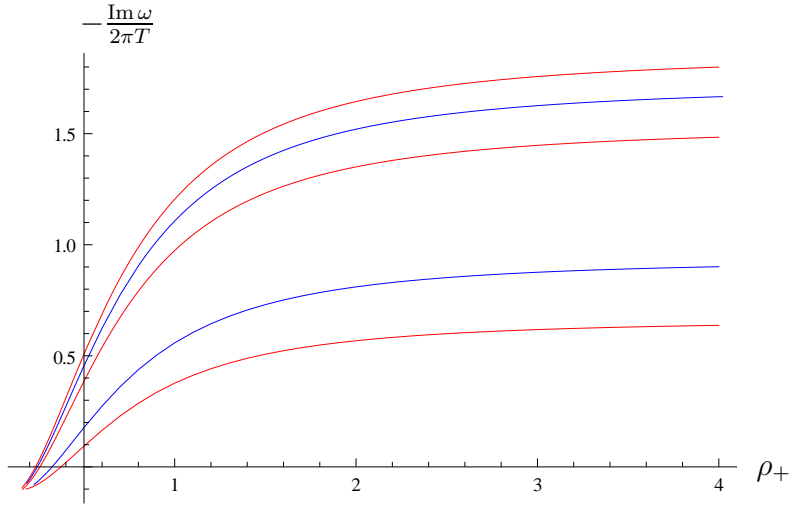}
\end{center}
  \caption{The dependence of $g=-\Im(\w)$ as a function of a black hole size 
$\r_+$ and temperature $T=\frac{2\r_+^2+1}{2\pi\r_+}$ in KW model for $U(1)_R$ charged/neutral quasinormal modes (red/blue) with $T^{1,1}$ eigenvalues $\l_{j,\ell,r}$: 
$(j,\ell,r)=\{(\ft 12,\ft 12, 1),(1,1,2),(\ft 32,\ft 12,1),(1,0,0),(1,1,0)\}$. $g$ increases with $\l_{j,\ell,r}$.
Black holes with $g<0$ are unstable with respect to condensation of 
these fluctuations.} \label{figure1}
\end{figure}

We now consider a simple application of \eqref{master} in the context of 
the holographic Klebanov-Witten (KW) model \cite{Klebanov:1999tb}. In this case $\calv_5$ is  $T^{1,1}$
coset manifold. Properties of the Laplacian on $T^{1,1}$ were extensively studied in \cite{Gubser:1998vd,Ceresole:1999ht,Ceresole:1999rq}. 
The eigenvalues are completely determined by a pair of $SU(2)$ spins $\{j,\ell\} \in \frac 12 \zet$ 
and a $U(1)_R$ $R$-symmetry charge $r\in \zet$ as
follows  
\begin{equation}
\l=\l_{j,\ell,r}= 6 \biggl(\
j(j+1)+\ell(\ell+1)-\frac 18 r^2
 \ \biggr)\,.
\eqlabel{lt11}
\end{equation}
A triplet $\{j,\ell,r\}$ is constraint so that both $2j$ and $2\ell$ have the same parity, and  
\begin{equation}
r\le \min\{2j,2\ell\}\,.
\eqlabel{rcondt}
\end{equation}
Note that the lowest non-vanishing eigenvalue on $T^{1,1}$ is 
\begin{equation}
\l_{min}=\l_{\ft 12,\ft 12, 1}=\frac{33}{4}\,.
\eqlabel{minl}
\end{equation}
The spectrum of the low-lying quasinormal modes of $AdS_5$ black holes in KW holography is shown on figure~\ref{figure1}.
The red curves correspond to states carrying $U(1)_R$ charge, and the blue curves represent neutral states. 

\begin{figure}[t]
\begin{center}
\psfrag{y}{{$\r_{+,crit}^2$}}
\psfrag{x}{{$\l$}}
\includegraphics[width=4in]{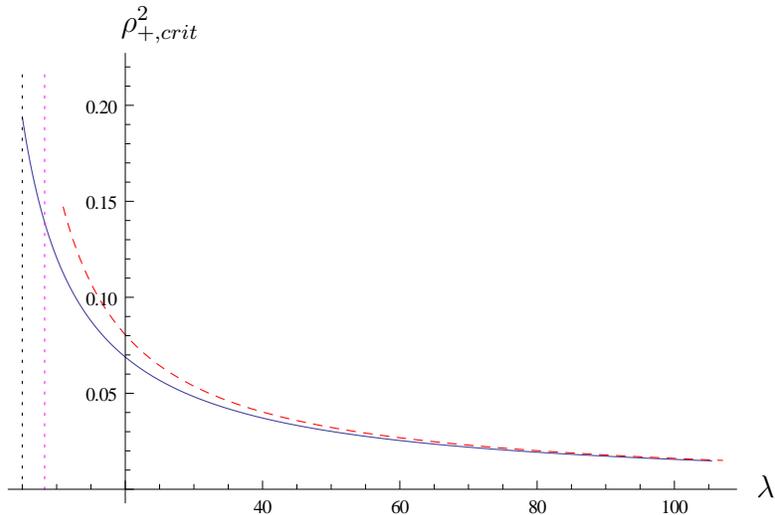}
\end{center}
  \caption{Critical size of the $AdS_5$ black hole $\r_{+,crit}^2$ below which a quasinormal mode with an eigenvalue $\l$ on $\calv_5$ becomes unstable.
The dashed red line is a large-$\l$ asymptotic \eqref{largel}. The black and 
the magenta vertical dotted lines  correspond to the onset of the 
instability (the smallest non-vanishing value of $\l$ on $\calv_5$) 
for $\calv_5=S^5$ and $\calv_5=T^{1,1}$ correspondingly.} \label{figure2}
\end{figure}

The solid blue curve in figure~\ref{figure2} presents the threshold value $\r_{+,crit}^2$ of the Gregory-Laflamme instability, corresponding to $\w=0$, for the 
$AdS_5\times \calv_5$ quasinormal mode with the $\calv_5$ eigenvalue $\lambda$. The dashed red line is the 
large-$\l$ asymptotic, see \cite{Buchel:2015gxa},
\begin{equation}
\r_{+,crit}^2=\frac{1.61015}{\l}+\calo(\l^{-2})\,.
\eqlabel{largel}
\end{equation}
Notice that larger values of $\l$ result in smaller threshold values of $\r_{+,crit}^2$. Thus, the onset of the instability of smeared $AdS_5\times \calv_5$ 
black holes is determined by the smallest non-vanishing eigenvalue $\l$ of the scalar Laplacian on $\calv_5$.


\section*{Acknowledgments}
 Research at Perimeter
Institute is supported by the Government of Canada through Industry
Canada and by the Province of Ontario through the Ministry of
Research \& Innovation. This work was further supported by
NSERC through the Discovery Grants program.

\end{document}